\documentclass[10pt, conference]{IEEEtran}
\IEEEoverridecommandlockouts
\usepackage{cite}
\usepackage{amsmath,amssymb,amsfonts}
\usepackage{algorithmic}
\usepackage{graphicx}
\usepackage{textcomp}
\usepackage{xcolor}
\def\BibTeX{{\rm B\kern-.05em{\sc i\kern-.025em b}\kern-.08em
    T\kern-.1667em\lower.7ex\hbox{E}\kern-.125emX}}

\usepackage{tikz}
\usepackage{pgfplots}
\usetikzlibrary{spy, arrows, patterns}
\pgfdeclarelayer{bg}    % declare background layer
\pgfsetlayers{bg,main}  % set the order of the layers (main is the standard layer)

\usepackage{bm}
\usepackage{float}
\usepackage{caption}
\usepackage{multirow}
\usepackage{booktabs}
\usepackage{subcaption}          % for subfigure
\newcommand{\img}{\mathbf{x}}
\newcommand{\sysout}{\hat{\img}}
\newcommand{\latent}{\mathbf{y}}
\newcommand{\qlatent}{\hat{\latent}}
\newcommand{\synthparam}{\bm{\theta}}
\newcommand{\armparam}{\bm{\psi}}
\newcommand{\upparam}{\bm{\upsilon}}
\newcommand{\synth}{f_{\synthparam}}
\newcommand{\arm}{f_{\armparam}}
\newcommand{\upsampling}{f_{\upparam}}
\newcommand{\analysisparam}{\bm{\alpha}}
\newcommand{\analysis}{f_{\analysisparam}}
\definecolor{redsynthesis}{HTML}{E76F51}
\definecolor{greenarm}{HTML}{2CA395}
\definecolor{purpleupsampling}{HTML}{6600CC}
\definecolor{bluelatent}{HTML}{829CBC}
\definecolor{orangenonoverfitted}{HTML}{FF9933}
\begin{document}

\title{Overfitted image coding at reduced complexity}

\author{\IEEEauthorblockN{Théophile Blard, Théo Ladune, Pierrick Philippe, Gordon Clare}
    \IEEEauthorblockA{\textit{Orange Innovation}, France \\ \texttt{\small{firstname.lastname@orange.com}}}
    \and
    \IEEEauthorblockN{Xiaoran Jiang, Olivier Déforges}
    \IEEEauthorblockA{\textit{IETR}, France \\ \texttt{\small{firstname.lastname@insa-rennes.fr}}}
    % \and
    % \IEEEauthorblockN{Théophile Blard}
    % \IEEEauthorblockA{\textit{Orange Innovation}, France \\ myemail@orange.com}
    % \and
    % \IEEEauthorblockN{Gordon Clare}
    % \IEEEauthorblockA{\textit{Orange Innovation}, France \\ myemail@orange.com}

}

\maketitle

\begin{abstract}
    Overfitted image codecs offer compelling compression performance and low
    decoder complexity, through the overfitting of a lightweight decoder for
    each image. Such codecs include Cool-chic, which presents image coding
    performance on par with VVC while requiring around 2000 multiplications per
    decoded pixel. This paper proposes to decrease Cool-chic encoding and
    decoding complexity. The encoding complexity is reduced by shortening
    Cool-chic training, up to the point where no overfitting is performed at all.
    It is also shown that a tiny neural decoder with 300 multiplications per
    pixel still outperforms HEVC. A near real-time CPU implementation of this
    decoder is made available at \texttt{\small{https://orange-opensource.github.io/Cool-Chic/}}.
\end{abstract}

\section{Introduction \& Related Works}

\begin{figure}[h]
    \centering
    \includegraphics[width=\linewidth]{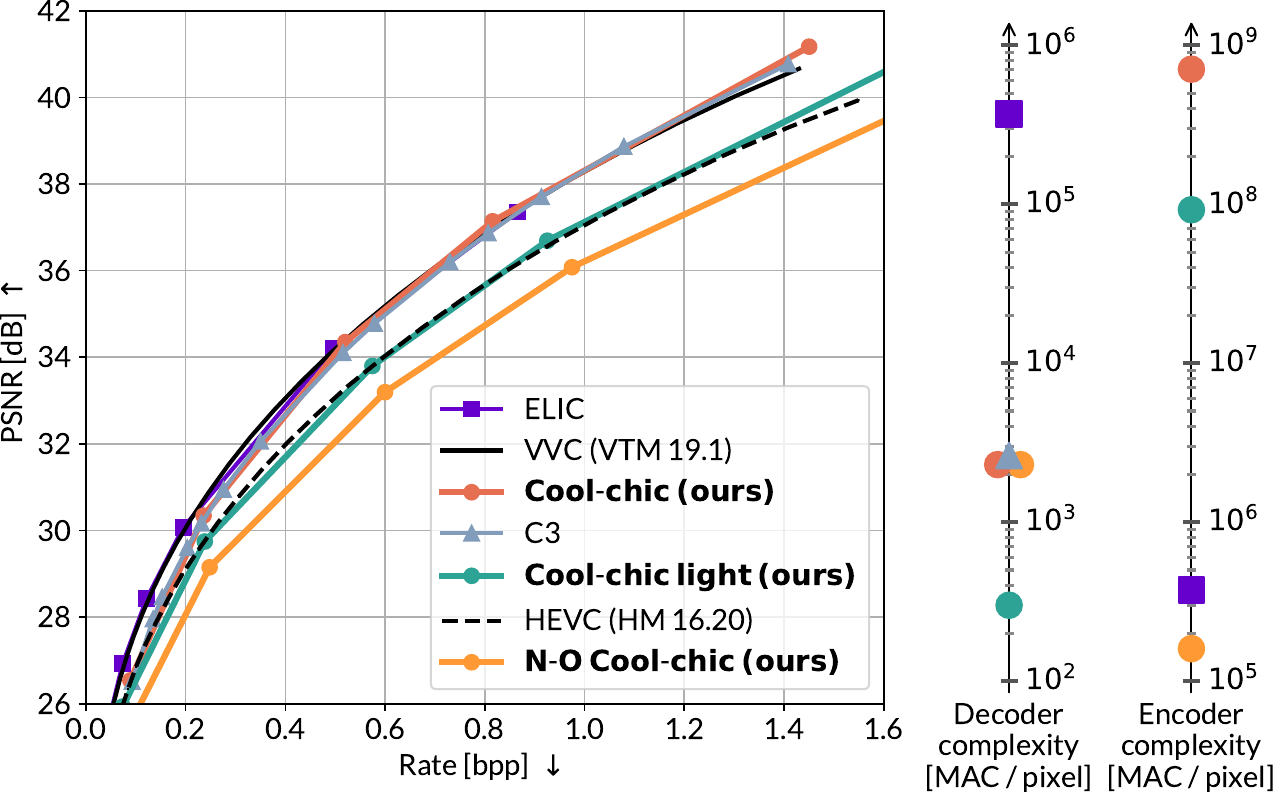}
    \caption{Rate-distortion performance on the Kodak \cite{kodak} dataset. MAC
    is for multiplication-accumulation. ELIC \cite{elic}, C3 \cite{c3}.}
    \label{fig:rd-curve-kodak}
\end{figure}

Autoencoder-based codecs (ELIC \cite{elic}, MLIC++ \cite{mlicpp}) offer
state-of-the-art compression results, outperforming conventional codecs
(H.265/HEVC \cite{hevc}, H.266/VVC \cite{vvc}). During training, autoencoder
parameters are optimized following an average rate-distortion cost computed on a
dataset supposed representative of all possible images. Once the training stage
is completed, parameters are frozen and the autoencoder relies on
\textit{generalization} to compress unseen images. Generalization requires
complex networks with many parameters, making the decoding prohibitively complex
with more than 100~000 multiplications per pixel (Fig.
\ref{fig:rd-curve-kodak}). This decoding complexity might hinder
autoencoder-based codec adoption especially when decoding happens on low-power
devices such as smartphones.
\newline

Overfitted codecs (Cool-chic \cite{coolchic1, coolchic2}, C3 \cite{c3}) have
emerged as an alternative paradigm to autoencoders. To compress an image,
overfitted codecs learn (overfit) a lightweight neural decoder and latent
representation tailored for this image. The decoder parameters are then conveyed
alongside the latents so that the receiver can reconstruct the image. Since
overfitted codecs do not rely on generalization, they offer compelling image
coding performance on par with VVC with a decoder complexity of 2000
multiplications per pixel (Fig. \ref{fig:rd-curve-kodak}). As for conventional
codecs, this is obtained through an expensive encoding process, during which the
decoder and latents are optimized according to the image rate-distortion cost.
\newline

This paper studies the effects of further simplification of overfitted codecs.
For these experiments, the open-source Cool-chic codec (which now implements
refinements from C3 \cite{c3}) is used \cite{coolchic-repository}. Our
contributions are as follows. The complexity-performance continuum of Cool-chic
is explored by reducing both the encoder and decoder complexity. It is shown
that a simple neural decoder with 300 MAC / pixel can outperform HEVC. A fast
CPU implementation of the decoder is demonstrated, showing that practical
decoding speed is possible even with the auto-regressive probability model
\cite{anti-arm-1,anti-arm-2}. A non-overfitted encoding is presented to address
cases where the encoding complexity constraint is paramount.

\begin{figure*}[t]
    \centering
    \includegraphics[width=\linewidth]{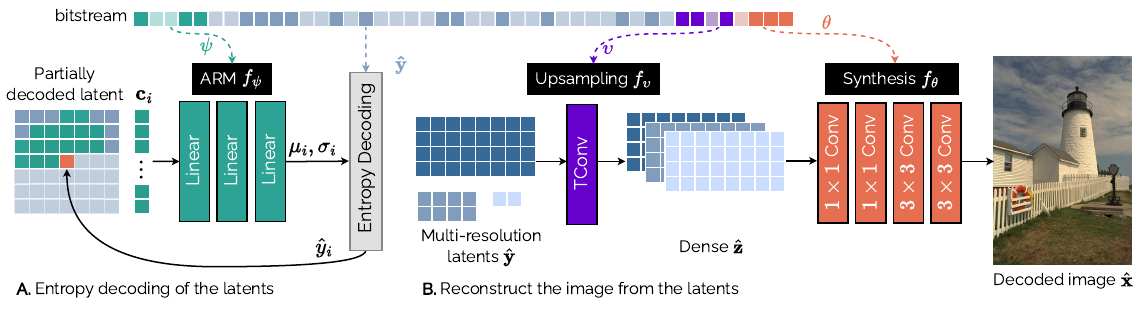}
    \caption{Cool-chic decoding. ARM: Auto-Regressive Model.}
    \label{fig:cool-chic-decoder}
\end{figure*}

\section{Background: Cool-chic}
\label{sec:background}

This section presents the Cool-chic encoding and decoding process. Note that
Cool-chic and C3 are both very similar, with a slightly different training
process. More details are available in previous work \cite{coolchic1, coolchic2,
c3}.
\newline

\textbf{Decoding.} Figure \ref{fig:cool-chic-decoder} presents the decoding
process of Cool-chic. It is composed of three main elements. \textit{i)} $L$
latent grids $\qlatent_l$ with different resolutions: $\qlatent = \{
\qlatent_{l} \in \mathbb{Z}^{H / 2^l \times W / 2^l},\ l = 0, \ldots, L - 1\}$.
\textit{ii)} The latent grids are transmitted using an entropy coding algorithm,
driven by an auto-regressive probability model $p_{\armparam}$ (ARM). This ARM
represents the distribution of one latent value conditioned on neighbouring
values and is implemented as a MLP $\arm$. \textit{iii)} The linear upsampling
$\upsampling$ and synthesis $\synth$ networks upsample the latents to a
dense representation and synthesize the decoded image $\sysout$.
\newline

\textbf{Encoding.} Cool-chic encodes an image $\img$ by simultaneously learning
the latent and the different neural networks according to the image
rate-distortion cost:
\begin{align}
    \qlatent^*, \armparam^*, \upparam^*, \synthparam^*
     & = \operatorname*{argmin}_{\qlatent, \armparam, \upparam, \synthparam} \mathrm{D}(\img, \sysout) + \lambda \mathrm{R}(\qlatent) \label{eq:coolchic-encoding}                        \\
     & = \operatorname*{argmin}_{\qlatent, \armparam, \upparam, \synthparam} || \img - \synth(\upsampling(\qlatent)) ||^2 - \lambda \log_2 p_{\armparam} \left(\qlatent\right). \nonumber
\end{align}
The Lagrange multiplier $\lambda \in \mathbb{R}$ balances the rate $\mathrm{R}$
and the distortion $\mathrm{D}$, here the mean-squared error. The discrete
latents $\qlatent = \mathrm{Q}(\latent)$ are optimized through the continuous
version $\latent$, using the method proposed in C3 \cite{c3} to obtain a
differentiable proxy for the quantization $\mathrm{Q}$. After the encoding,
neural network parameters are quantized and entropy coded based on an ad-hoc
probability model \cite{coolchic1} while the latents are entropy coded with a
range coder driven by the probability model $p_{\armparam}$.

\section{Towards a fast and lightweight decoder}

Current state-of-the-art overfitted codecs achieve compression performance
comparable to VVC with around 3~000 multiplications per decoded pixel \cite{c3}.
This section proposes a lighter and faster Cool-chic decoder, either through
complexity reduction or better implementation.

\subsection{Decoder complexity-performance continuum}

\begin{figure*}[h]
    \begin{subfigure}{0.49\linewidth}
        \centering
        \includegraphics[width=0.95\linewidth]{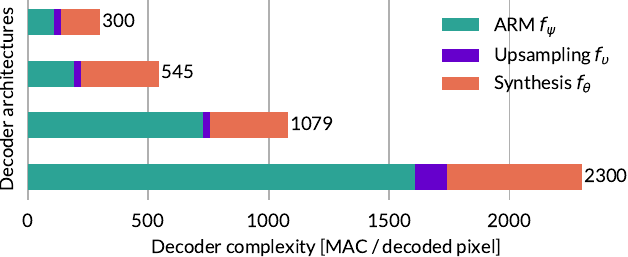}
        \caption{Detailed decoder complexity.}
        \label{fig:decoder-complexity}
    \end{subfigure}
    \begin{subfigure}{0.49\linewidth}
        \centering
        \includegraphics[width=0.95\linewidth]{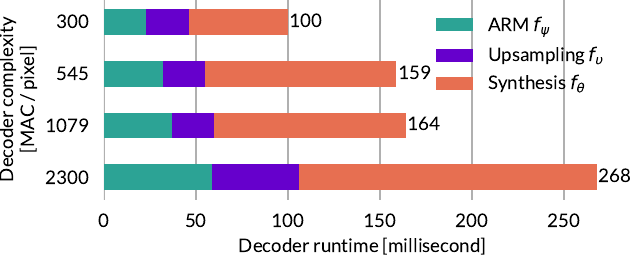}
        \caption{Average CPU (AMD EPYC 7282) decoding time on Kodak.}
        \label{fig:decoder-runtime-kodak}
    \end{subfigure}
    \caption{Complexity and runtime for different Cool-chic decoding architectures.}
\end{figure*}

\textbf{Lighter decoders.} Cool-chic decoding complexity is varied by changing
the architecture of its different neural networks. Several complexity tradeoffs
were evaluated and the best ones were retained. Table
\ref{tab:decoder-architecture} presents the selected architectures, ranging from
300 to 2300 MAC (multiplication-accumulation) per decoded pixel. Figure
\ref{fig:decoder-complexity} details the share of each network in the overall
decoding complexity, highlighting that most of the complexity comes from the ARM
network. Empirically, it is found that the ARM is the module which most benefits
from additional complexity.
\newline

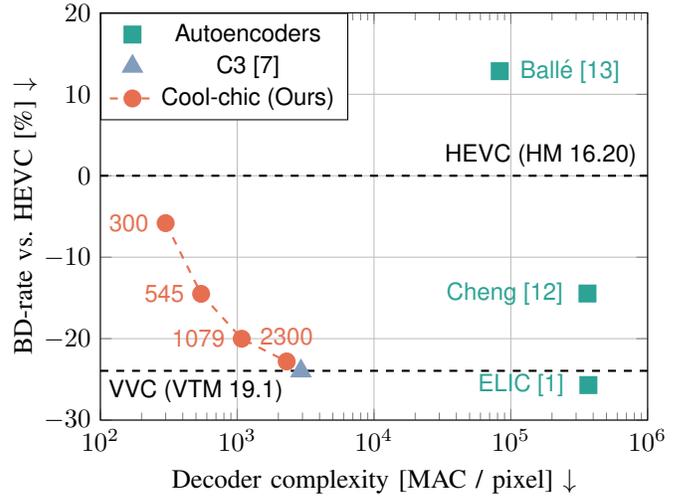
\begin{figure}[t]
        \pgfdeclarelayer{bg}    % declare background layer
        \pgfsetlayers{bg,main}  % set the order of the layers (main is the standard layer)
        \centering
        \begin{tikzpicture}
            \begin{semilogxaxis}[
                    grid= major,
                    width=\linewidth,
                    height=7cm,
                    xlabel = {Decoder complexity [MAC / pixel] $\downarrow$},
                    ylabel = {BD-rate vs. HEVC [\%] $\downarrow$} ,
                    xmin = 100, xmax = 1000000, xlabel near ticks, minor x tick num=0,
                    ymin = -30, ymax = 20, ylabel near ticks, minor y tick num=0, ytick distance={10},
                    title style={yshift=-0.75ex},
                    ylabel shift=-0.15cm,
                    legend style={at={(0.,1.)}, anchor= north west},
                ]

                % ------------------ Draw the different points ------------------ %
                \addplot[thick, greenarm, only marks, mark=square*, mark size=3pt] coordinates {
                        (83000 , 12.837)      % Balle 18 hyperprior
                        (363350, -14.459)     % Cheng 2020
                        (368997, -25.7)       % ELIC

                    };
                \addlegendentry{\sf \small Autoencoders}

                \addplot[thick, bluelatent, only marks, mark=triangle*, mark size=4pt] coordinates {
                        (2925 , -23.98)      % C3 CLIC

                    };
                \addlegendentry{\sf \small  C3 \cite{c3}}

                \addplot[thick, dashed, redsynthesis, mark=*, mark size=3pt, mark options={solid}] coordinates {
                        (300, -5.82)
                        (545, -14.50)
                        (1079, -20.01)
                        (2300, -22.8)
                    };
                \addlegendentry{\sf \small Cool-chic (Ours)}

                \draw[dashed, thick, black] (axis cs:100, 0) -- (axis cs:1000000, 0);
                \node [black, above, xshift=0.4cm] at (axis cs:100000, 0){\sf \small HEVC (HM 16.20)};

                \draw[dashed, thick, black] (axis cs:100, -23.95) -- (axis cs:1000000, -23.95);
                \node [black, below, xshift=0.4cm] at (axis cs:300, -23.95){\sf \small VVC (VTM 19.1)};

                % ----------------------- Write the names ----------------------- %
                \node [redsynthesis, left, xshift=-0.10cm] at (axis cs:300, -5.82){\sf \small 300};
                \node [redsynthesis, left, xshift=-0.10cm] at (axis cs:545, -14.5){\sf \small 545};
                \node [redsynthesis, left, xshift=-0.10cm] at (axis cs:1079, -20.01){\sf \small 1079};
                \node [redsynthesis, above, yshift=0.10cm] at (axis cs:2300, -22.8){\sf \small 2300};
                \node [greenarm, right, xshift=0.15cm] at (axis cs:83000  , 12.837){\sf \small Ballé \cite{balle}};
                \node [greenarm, left, xshift=-0.2cm] at (axis cs:363350 , -14.459){\sf \small Cheng \cite{ChengSTK20}};
                \node [greenarm, left, xshift=-0.2cm] at (axis cs:368997 , -25.7){\sf \small ELIC \cite{elic}};
                % ----------------------- Write the names ----------------------- %
            \end{semilogxaxis}
        \end{tikzpicture}
    %     \caption{CLIC 2020 professional validation set \cite{clic20pro}}
    %     \label{fig:decoder-complexity-bdrate-clic}
    % \end{subfigure}
    \caption{Rate savings versus HEVC on CLIC 2020 professional validation set
        \cite{clic20pro}. Negative results: less rate is required to get the
        same quality as HEVC.}
    \label{fig:decoder-complexity-bdrate}

\end{figure}

\textbf{Results.} The rate-distortion performance is presented in Fig.
\ref{fig:rd-curve-kodak} and \ref{fig:decoder-complexity-bdrate}. In its
lightest setting (300 MAC), Cool-chic outperforms the well-established
conventional codec HEVC and its most complex configuration (2300 MAC) is
competitive with VVC. Furthermore, Cool-chic also outperforms autoencoder-based
codecs such as Cheng \textit{et al.} \cite{ChengSTK20} while being approximately
1000 times less complex at the decoder-side. These are compelling results, as no
previous learned codecs have been able to outperform HEVC with only 300
multiplications per decoded pixel.

\begin{table*}[h]
    % \captionsetup{skip=10pt} % Increase the spacing between the table and the caption
    \centering
    \small
    \begin{tabular}{cc||c | c | c}
        \multicolumn{2}{c||}{Decoder} & \multirow{2}{*}{ARM $\arm$} & \multirow{2}{*}{Upsampling $\upsampling$} & \multirow{2}{*}{Synthesis $\synth$}                                                      \\
        MAC/pixel                     & Param                       &                                           &                                     &                                                    \\
        \midrule
        300                           & 281                         & 8-8-LinR, 8-2-Lin                         & 1-1-TConv-4                         & 7-8-Conv-1, 8-3-Conv-1, 3-3-ConvR-3                \\
        545                           & 525                         & 8-8-LinR, 8-8-LinR, 8-2-Lin               & 1-1-TConv-4                         & 7-16-Conv-1, 16-3-Conv-1, 3-3-ConvR-3, 3-3-ConvR-3 \\
        1079                          & 941                         & 16-16-LinR, 16-16-LinR, 16-2-Lin          & 1-1-TConv-4                         & 7-16-Conv-1, 16-3-Conv-1, 3-3-ConvR-3, 3-3-ConvR-3 \\
        2300                          & 1925                        & 24-24-LinR, 24,24-LinR, 24-2-Lin          & 1-1-TConv-8                         & 7-40-Conv-1, 40-3-Conv-1, 3-3-ConvR-3, 3-3-ConvR-3 \\
    \end{tabular}
    \caption{Decoder architectures. Layers are denoted as
        \textit{InFeat-OutFeat-Layer-Kernel}. \textit{Lin}, \textit{TConv} and
        \textit{Conv} respectively denotes linear, convolution or transpose
        convolution layer and \textit{R} stands for residual. Transpose
        convolutions use a stride of 2. All layers are followed by a ReLU
        linearity except for the last one of each module.}
    \label{tab:decoder-architecture}
\end{table*}

\subsection{Fast decoder implementation}

\textbf{Parallelization.} Previous work \cite{anti-arm-1,anti-arm-2} states that
despite its low number of multiplications, Cool-chic ARM is intrinsically slow
due to its sequential nature. Sequentiality is usually frowned upon for
autoencoder-based codecs, since they primarily target a GPU to offer fast enough
decoding. Here, Cool-chic low decoding complexity allows it to run on
non-dedicated hardware \textit{e.g.} CPU where parallelization is not as
important. To assess more accurately the practicality of Cool-chic, a faster
CPU-only C implementation of the decoder is provided on the Cool-chic repository
\cite{coolchic-repository}.
\newline

\textbf{Fast C implementation.} Figure \ref{fig:decoder-runtime-kodak} presents
the decoder runtime for different complexities. Even for the most complex
decoder, the ARM requires only 50 milliseconds, making Cool-chic CPU-only
decoding practical. To achieve efficient entropy decoding, the range coder is
switched to the binary arithmetic coder of HEVC. Note that most of the
optimisations concern the ARM \textit{e.g.} using AVX2 instructions if
available. Consequently, the runtime (Fig. \ref{fig:decoder-runtime-kodak}) is
not proportional to the complexity (Fig. \ref{fig:decoder-complexity}): while
the ARM represents up to 75~\% of the multiplications, it never accounts for
more than 20~\% of the runtime. Since the ARM is arguably the less
implementation-friendly module, this hints that both the upsampling and the
synthesis could be made even faster. In particular, the current implementation
is single core even though multi-threading would be beneficial for these two
modules.

% \begin{figure}[h]
%     \centering
%     \includegraphics[width=0.9\linewidth]{figures/decoder_runtime_kodak.pdf}
%     \caption{Average CPU decoding time on Kodak. CPU model is AMD EPYC 7282, 2.8 GHz.}
%     \label{fig:decoder-runtime-kodak}
% \end{figure}

\section{Reducing the encoding complexity}

Cool-chic is designed to reduce, first and foremost, the decoding complexity.
Nevertheless, use-cases arise where the encoding complexity becomes the main
constraint. This section studies the impact of a shorter training and proposes a
solution for situations where the encoding complexity constraint is paramount: a
non-overfitted (N-O) Cool-chic.

\subsection{Shorter training}

\textbf{Experimental setup.} As with conventional codecs, Cool-chic relies on
encoding-time rate-distortion optimization to achieve compelling compression
while maintaining a low decoder complexity. Reducing the training (\textit{i.e.}
encoding) time directly reduces the encoding complexity at the expense of the
compression performance. Cool-chic encoding complexity is measured in MAC per
encoded pixel. One training iteration consists in one forward pass and one
backward pass. During the backward pass, each layer computes 2 quantities: the
gradient of the layer output with regards to layer weights (similar complexity
than the forward pass), and the gradient of the layer output with respect to the
layer input. Consequently, the complexity of the backward pass is approximated
to twice the MAC of the forward pass \cite{baydin2018automatic}. Let
$\kappa_{dec}$ denotes the decoder complexity in MAC / pixel, the encoding
complexity $\kappa_{enc}$ is
\begin{equation}
    \kappa_{enc} = 3N \times \kappa_{dec},\ \text{with } N \text{ the number of iterations.}
\end{equation}

% Furthermore, we assume that the encoding is performed on a device with enough
% computational ressources to train up to 8 independent Cool-chic decoders
% simultaneously. Since learning 8 decoders and picking up the best one takes as
% long as learning a single one, we only account for the training complexity
% of the selected decoder.
% \newline

\textbf{Results.} Figure \ref{fig:encoder-complexity-bdrate-clic} presents the
compression performance as a function of the encoding complexity. The best
results are obtained with an encoding complexity close to $10^9$ MAC / pixel,
which is orders of magnitude more complex than for autoencoder-based codecs.
Cool-chic encoding complexity can be reduced by one order of magnitude
(\textit{i.e.} to $10^8$ MAC / pixel) with a slight BD-rate increase (less than
10~\%). The encoding can still be made simpler ($10^7$ MAC / pixel) while
maintaining performance on par with HEVC. It is worth noting than for lower
encoding complexity, having a lighter decoder (\textit{e.g.} 300 MAC / pixel) is
better than a more complex decoder. Even at this point, the remaining encoding
complexity of Cool-chic is still significantly higher than autoencoders.
Reducing this complexity further requires a paradigm shift presented in the
following section.
\newline

\subsection{Non-overfitted Cool-chic}

\begin{figure*}[h]
    \centering
    \includegraphics[width=\linewidth]{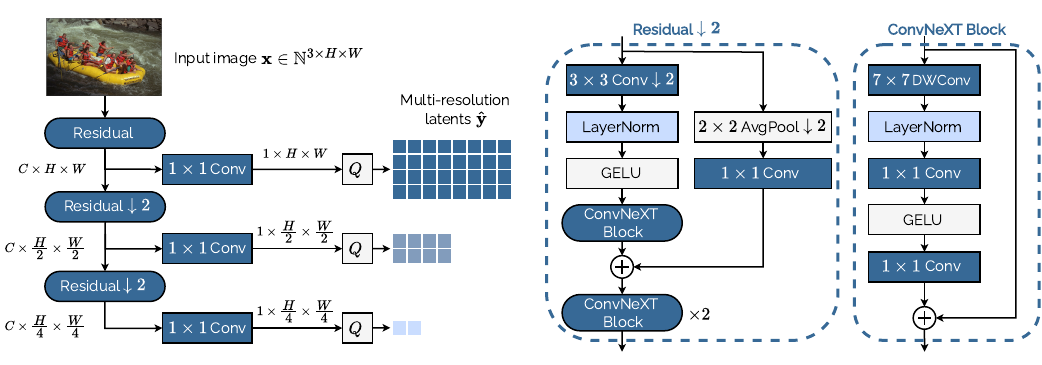}
    \caption{Proposed analysis transform. \textit{DWConv} stands for depth-wise convolution and $\downarrow$ denotes the stride.}
    \label{fig:cool-chic-analysis}
\end{figure*}

\textbf{Analysis transform. }In order to reach even shorter encoding times, the
overfitting process can be completely bypassed by training a non-overfitted
(N-O) Cool-chic, sharing identical parameters for all images. Taking inspiration
from autoencoders, encoding an image with N-O Cool-chic relies on an additional
analysis transform $\analysis$ generating a Cool-chic-compatible latent
representation $\qlatent$ from the input image $\img$:
\begin{equation}
    \qlatent = \analysis(\img).
\end{equation}

\textbf{Analysis architecture. }The proposed analysis transform is depicted in Fig
\ref{fig:cool-chic-analysis}. To produce a set of $L = 7$ hierarchical latent
grids, a series of $L$ residual blocks progressively downsample the input image
$\img$ and extract relevant features for the different resolutions. After each
downsampling step, a simple $1\times 1$ convolution merges the $C = 64$ features
into the $l$-th latent $\qlatent_l$. The proposed analysis hence produces a few latent
grids with hierarchical resolutions, unlike autoencoder analysis which computes
hundred of small-resolution features.
\newline

\textbf{Residual blocks. }To increase the receptive field while maintaining low
complexity, ConvNeXt blocks \cite{liu2022convnet} are adopted upon conventional
residual blocks \cite{he2016deep}. Downsampling with residual block is achieved
by complementing the identity branch with a stride-2 $2\times 2$ average pooling
and a $1\times 1$ convolution as in \cite{he2019bag}. The first residual block does
not downsample so the pooling layer is removed. Following ELIC \cite{elic},
additional residual blocks are stacked after each downsampling for more
expressivity.
\newline

\textbf{Decoder. } The decoder parameters $(\armparam, \upparam, \synthparam)$
are learned alongside the analysis transform. The architecture used for this
experiments is $\kappa_{dec}=2300$ from Table \ref{tab:decoder-architecture}
which gives the best rate-distortion performance. Note that the network
parameters no longer have to be sent alongside the latent representation since
they are shared for all images.
\newline

\textbf{Training. }The proposed N-O Cool-chic aims to obtain optimal parameters,
generalizable to all possible images. To this end, it is trained with $256\times
256$ patches of randomly cropped images from the CLIC 2019 training set
\cite{clic19}. Since the latent representation $\qlatent$ is now obtained from
the analysis transform $\analysis$, eq. \eqref{eq:coolchic-encoding} becomes
\begin{align}
    \analysisparam^*, & \armparam^*, \upparam^*, \synthparam^*
    = \operatorname*{argmin}_{\analysisparam, \armparam, \upparam, \synthparam} \mathbb{E}_{\img} [\mathrm{D}(\img, \sysout) + \lambda \mathrm{R}(\qlatent)] \label{eq:non-overfitted-coolchic-encoding}                                         \\
                    & = \operatorname*{argmin}_{\analysisparam, \armparam, \upparam, \synthparam} \mathbb{E}_{\img} [ || \img - \synth(\upsampling(\analysis(\img))) ||^2 - \lambda \log_2 p_{\armparam} \left(\analysis(\img)\right)]. \nonumber
\end{align}
Independent training with different rate constraints is performed to cover a
wide range of rates, namely $\lambda = \{ 0.02, 0.004, 0.001, 0.0004, 0.0001
\}$. Adam algorithm \cite{kingma2014adam} is used, the learning rate starts from
$10^{-3}$ and is dynamically decayed by a patience mechanism. When it reaches
$10^{-6}$, the training is terminated. Latent variable quantization is similar
to the overfitted Cool-chic (section \ref{sec:background}).
\newline

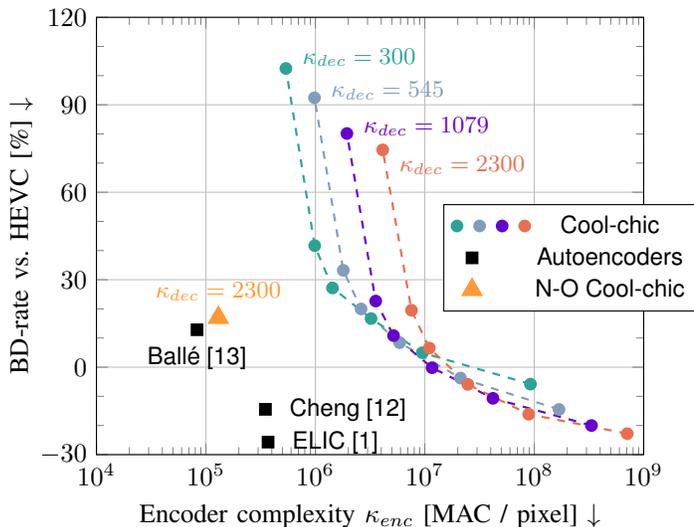
\begin{figure}[h]
    \pgfdeclarelayer{bg}    % declare background layer
    \pgfsetlayers{bg,main}  % set the order of the layers (main is the standard layer)
    \centering
    \begin{tikzpicture}
        \begin{semilogxaxis}[
                grid= major,
                width=\linewidth,
                height=7.4cm,
                xlabel = {Encoder complexity $\kappa_{enc}$ [MAC / pixel] $\downarrow$},
                ylabel = {BD-rate vs. HEVC [\%] $\downarrow$} ,
                xmin = 10000, xmax = 1000000000, xlabel near ticks, minor x tick num=0,
                ymin = -30, ymax = 120, ylabel near ticks, minor y tick num=2, ytick distance={30},
                title style={yshift=-0.75ex},
                ylabel shift=-0.15cm,
                legend style={at={(1.1,0.45)}, anchor= east},
                % define plot styles, for convenience
                cc300/.style={thick, greenarm, dashed, mark=*, mark options={solid}, mark size=2pt},
                cc545/.style={thick, bluelatent, dashed, mark=*, mark options={solid}, mark size=2pt},
                cc1079/.style={thick, purpleupsampling, dashed, mark=*, mark options={solid}, mark size=2pt},
                cc2300/.style={thick, redsynthesis, dashed, mark=*, mark options={solid}, mark size=2pt},
                % make new legend style
                combo legend/.style={
                        legend image code/.code={
                                \draw [/pgfplots/cc300] plot coordinates {(1mm,0cm)};
                                % \draw plot coordinates {(2.5mm,-3pt)} node {,};
                                \draw [/pgfplots/cc545] plot coordinates {(4mm,0cm)};
                                % \draw plot coordinates {(6.5mm,-3pt)} node {,};
                                \draw [/pgfplots/cc1079] plot coordinates {(7mm,0cm)};
                                % \draw plot coordinates {(10.5mm,-3pt)} node {,};
                                \draw [/pgfplots/cc2300] plot coordinates {(10mm,0cm)};
                            }
                    }
            ]

            % ------------------ Draw the different points ------------------ %

            \addplot[thick, black, dashed, mark=*, mark options={solid}, mark size=2pt] coordinates {
                };
            \addlegendentry{\sf \small Cool-chic}

            % adds new "fake plot" that is included in legend
            \addlegendimage{combo legend}

            \addplot[thick, greenarm, dashed, mark=*, mark options={solid}, mark size=2pt, forget plot] coordinates {
                    (92346150,	-5.82)
                    (9508200,	4.95)
                    (3229200,	16.67)
                    (1435200,	27.17)
                    (986700,	41.67)
                    (538200,	102.46)

                };

            \addplot[thick, bluelatent, dashed, mark=*, mark options={solid}, mark size=2pt, forget plot] coordinates {
                    (168415905	,	-14.50)
                    (21184905	,	-3.73)
                    (5889240,	8.41)
                    (2617440,	19.97)
                    (1799490,	33.21)
                    (981540,	92.39)
                };

            \addplot[thick, purpleupsampling, dashed, mark=*, mark options={solid}, mark size=2pt, forget plot] coordinates {
                    (333249150,	-20.01)
                    (41919150,	-10.68)
                    (11653200,	-0.19)
                    (5179200,	10.83)
                    (3560700,	22.68)
                    (1942200,	80.15)
                };

            \addplot[thick, redsynthesis, dashed, mark=*, mark options={solid}, mark size=2pt, forget plot] coordinates {
                    (707575350,		-22.80)
                    (89005350,		-16.14)
                    (24742800,		-5.91)
                    (10996800,		6.56)
                    (7560300,		19.52)
                    (4123800,		74.53)
                };

            \addplot[thick, black, only marks, mark=square*, mark size=2pt] coordinates {
                    (83000 , 12.84)      % Balle 18 hyperprior
                    (350000, -14.46)     % Cheng 2020
                    (370000, -25.7)       % ELIC

                };
            \addlegendentry{\sf \small Autoencoders}

            \addplot[thick, orangenonoverfitted, only marks, mark=triangle*, mark size=4pt] coordinates {
                    (130000 , 17.0)
                };
            \addlegendentry{\sf \small N-O Cool-chic}

            % \addplot[thick, bluelatent, only marks, mark=triangle*, mark size=4pt] coordinates {
            %     (2925 , -23.98)      % C3 CLIC

            % };
            % \addlegendentry{\sf \small Overfitted codecs}

            % \addplot[thick, dashed, redsynthesis, mark=*, mark size=4pt, mark options={solid}] coordinates {
            %     (300, -5.82)
            %     (545, -14.50)
            %     (1079, -20.01)
            %     (2300, -22.8)
            % };
            % \addlegendentry{\sf \small \textbf{Cool-chic (Ours)}}

            % \draw[dashed, thick, black] (axis cs:100, -23.95) -- (axis cs:1000000, -23.95);
            % \node [black, below, xshift=0.4cm] at (axis cs:300, -23.95){\sf \small VVC (VTM 19.1)};

            % % ----------------------- Write the names ----------------------- %
            % \node [redsynthesis, left, xshift=-0.10cm] at (axis cs:300, -5.82){\sf \small 300};
            % \node [redsynthesis, left, xshift=-0.10cm] at (axis cs:545, -14.5){\sf \small 545};
            % \node [redsynthesis, left, xshift=-0.10cm] at (axis cs:1079, -20.01){\sf \small 1079};
            % \node [redsynthesis, above, yshift=0.10cm] at (axis cs:2300, -22.8){\sf \small 2300};
            % \node [bluelatent, below, yshift=-0.05cm, xshift=0.4cm] at (axis cs:2925  , -23.98){\sf \small C3 \cite{c3}};
            \node [greenarm, right, xshift=0.10cm, yshift=0.15cm] at (axis cs:538200,	102.46){\sf \small $\kappa_{dec} = 300$};
            \node [bluelatent, right, xshift=0.10cm, yshift=0.1cm] at (axis cs:981540,	92.39){\sf \small $\kappa_{dec} = 545$};
            \node [purpleupsampling, right, xshift=0.10cm, yshift=0.1cm] at (axis cs:1942200,	80.15){\sf \small $\kappa_{dec} = 1079$};
            \node [redsynthesis, right, xshift=0.10cm, yshift=-0.2cm] at (axis cs:4123800,		74.53){\sf \small $\kappa_{dec} = 2300$};
            \node [orangenonoverfitted, above, yshift=0.1cm] at (axis cs:130000 , 17.0){\sf \small $\kappa_{dec} = 2300$};
            \node [black, below, yshift=-0.1cm] at (axis cs:83000  , 12.837){\sf \small Ballé \cite{balle}};
            \node [black, right, xshift=0.2cm] at (axis cs:350000 , -14.459){\sf \small Cheng \cite{ChengSTK20}};
            \node [black, right, xshift=0.2cm] at (axis cs:370000 , -25.7){\sf \small ELIC \cite{elic}};
            % % ----------------------- Write the names ----------------------- %
        \end{semilogxaxis}
    \end{tikzpicture}
    \caption{Rate savings versus HEVC (HM 16.20) as a function of the encoding
        complexity. Negative results: less rate is required to get the same
        quality than HEVC. CLIC 2020 professional validation set
        \cite{clic20pro}}
    \label{fig:encoder-complexity-bdrate-clic}
\end{figure}

\textbf{Results. }Figure \ref{fig:encoder-complexity-bdrate-clic} shows the
compression performance of N-O Cool-chic against the encoding complexity and
Fig. \ref{fig:rd-curve-kodak} presents the rate-distortion curves. N-O Cool-chic
encodes images in a single forward pass through the analysis. This reduces the
complexity by a factor of 1000 compared to Cool-chic, requiring only 160 kMAC
per pixel which is comparable to other autoencoder-based codecs. Beside the
reduction in MAC, encoding images with N-O Cool-chic is conceptually simpler
since there is no more optimization through gradient descent. However, this
reduces the performance significantly, falling behind state-of-the art codecs
(Cheng \cite{ChengSTK20} or ELIC \cite{elic}). Compared to Ballé \cite{balle},
N-O Cool-chic requires 3\% more rate to achieve the same quality, with similar
encoding complexity but with a decoder 20 times less complex.
\newline

\section{Future work \& discussion}

\textbf{On the benefits of overfitting. }Overfitted codecs such as Cool-chic are
inherently adaptive. Contrary to autoencoders which rely on a fixed set of
parameters, the decoder and latent representation are specifically tuned
to offer the best performance possible on each single image. This adaptivity
allows overfitted codecs to perform on par with state-of-the-art neural codecs
while maintaining a low decoding complexity. N-O Cool-chic uses an analysis
transform and a shared decoder, which makes its functioning similar to
autoencoders. The performance loss between N-O Cool-chic and Cool-chic
highlights the benefit of overfitting, although this gap could be mitigated with
an improved analysis transform.
\newline

\textbf{Cool-chic limitations. }The main drawback of current overfitted codecs
remains their encoding time. To obtain the best rate-distortion performance,
the encoding time can reach 30 minutes for a single $512\times768$ image. This paper
shows that it can be reduced by a factor 10 ($\simeq$ 3 minutes) to 100
($\simeq$ 20 seconds) without major performance decrease. While
convenient, this is not enough for use-cases where \textit{real-time} encoding
is required. To this end N-O Cool-chic is proposed, where the encoding becomes a
simple forward pass (less than 1 second) at the expense of worse rate-distortion
performance.
\newline

\textbf{Meta learning. } We believe that the performance gap between overfitted
and N-O Cool-chic would be greatly reduced by introducing a \textit{slightly
overfitted} version, inspired by meta-learned approaches such as COIN++
\cite{coinplusplus}. Carefully selecting a few parameters to be overfitted might
allow an encoding of images that is faster than training a whole decoder from scratch for
every image. Also, the decoder parameters and the latent representation
computed by the analysis could be fine-tuned image-wise as in
\cite{DBLP:conf/iccv/YangM23}.

\section{Conclusion}

This paper strives to make Cool-chic encoding and decoding faster. It is shown
that Cool-chic outperforms HEVC, even with a lightweight decoder requiring 300
multiplications per pixel. A fast implementation of Cool-chic decoder is
provided, requiring 100 milliseconds to decode an image on CPU. This hints that
fast neural decoding is possible, even without dedicated hardware. Then, the
performance-encoding complexity continuum is investigated. In particular, it is
found that encoding can be made almost 100 times faster while still being
competitive with HEVC. Finally, a non-overfitted (N-O) Cool-chic is designed for
use-cases with extreme constraints on the encoding complexity, reducing the
encoding complexity up to a factor of 1000.

\bibliographystyle{IEEEbib}
\bibliography{refs}

\end{document}